\documentclass[aps,prl,twocolumn,superscriptaddress,showpacs]{revtex4}
 \usepackage{graphicx}
\usepackage{afterpage} \usepackage{dcolumn} \usepackage{bm}

\begin{document}

\preprint{APS/123-QED}

%\draft

%\twocolumn[\hsize\textwidth\columnwidth\hsize\csname@twocolumnfalse\endcsname

\title{Far-infrared probe of superconductivity in SmO$_{1-x}$F$_{x}$FeAs}

\author{S.I. Mirzaei}\affiliation{D\'epartement de Physique de la Mati\`ere Condens\'ee, Universit\'e de Gen\`eve, Gen\`eve, Switzerland.}
\author{V. Guritanu}\affiliation{D\'epartement de Physique de la Mati\`ere Condens\'ee, Universit\'e de Gen\`eve, Gen\`eve, Switzerland.}
\author{A. B. Kuzmenko}\affiliation{D\'epartement de Physique de la Mati\`ere Condens\'ee, Universit\'e de Gen\`eve, Gen\`eve, Switzerland.}
\author{C. Senatore}\affiliation{D\'epartement de Physique de la Mati\`ere Condens\'ee, Universit\'e de Gen\`eve, Gen\`eve, Switzerland.}
\author{D. van der Marel}\affiliation{D\'epartement de Physique de la Mati\`ere Condens\'ee, Universit\'e de Gen\`eve, Gen\`eve, Switzerland.}
\author{G. Wu}\affiliation{Hefei National Laboratory for Physical Science at Microscale and Department of Physics, University of Science and Technology of China, China\\}
\author{R. H. Liu}\affiliation{Hefei National Laboratory for Physical Science at Microscale and Department of Physics, University of Science and Technology of China, China\\}
\author{X. H. Chen}\affiliation{Hefei National Laboratory for Physical Science at Microscale and Department of Physics, University of Science and Technology of China, China\\}
\date{\today}


\begin{abstract}
We report far-infrared reflectance measurements on
polycrystalline superconducting samples of
SmO$_{1-x}$F$_{x}$FeAs ($x$ = 0.12, 0.15 and 0.2). We clearly
observe superconductivity induced changes of reflectivity in a
broad range of energies, which resembles earlier optical
measurements on high $T_{c}$ cuprates. The
superconducting-to-normal reflectivity ratio $R_{s}/R_{n}$
grows for the photon energies below 18 meV and shows a
complicated structure due to the presence of a strong
infrared-active phonon at about 10 meV.
\end{abstract}

\maketitle

The discovery of high Tc superconductivity in the iron pnictides\cite{KamiharaJACS08} has provoked a vivid debate concerning the microscopic origin of superconductivity in these 3d transition metal compounds. Motivated by the fact that electron correlation effects play a dominant role in the physical properties of 3d transition metals, it is believed that the pairing interaction
in the iron pnictides is mediated by coupling to spin-fluctuations as opposed to
phonons \cite{MazinCM08,BoeriCM08,DaiCM08,RaghuCM08,MaCM08}.

Far infrared optical spectroscopy is an important probe of the
superconducting state relative deep inside the sample due to the fact that the skin depth at these frequencies is on the order of 100 nanometer. For superconductors with a T$_c$ of 10 K  or more the gap is in range of modern infrared spectrometers\cite{TinkhamBook}. Also the other parameters, such as the scattering rate, the plasma frequency, and the intrinsic c-axis Josephson plasma-frequencies of strongly anisotropic high T$_c$ superconductors usually fall inside the range of infrared and optical spectrometers\cite{tsvetkov98}.
We measured the reflectivity of SmO$_{1-x}$F$_{x}$FeAs polycrystals above and below the
superconducting transition. The samples were synthesized by
conventional solid state reaction as described elsewhere
\cite{ChenCM08Synthesis,SenatoreCM08}. The grain size is
estimated to be maximum 3 $\mu$m\cite{SenatoreCM08}. Because of a
relatively large porosity, we have chosen to use unpolished
surfaces for infrared measurements.

\begin{figure}[thb]
   \centerline{\includegraphics[width=\linewidth, clip=true]{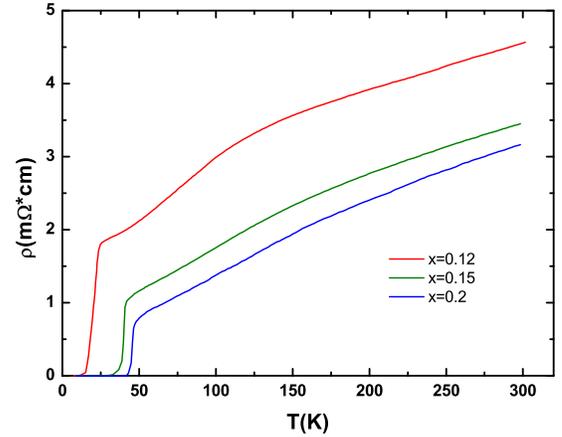}}
   \caption{The resistivity of the SmFeAsO$_{1-x}$F$_{x}$ samples with different doping as a function of temperature.}
   \label{FigResist}
\end{figure}

\begin{figure}[thb]
   \centerline{\includegraphics[width=\linewidth, clip=true]{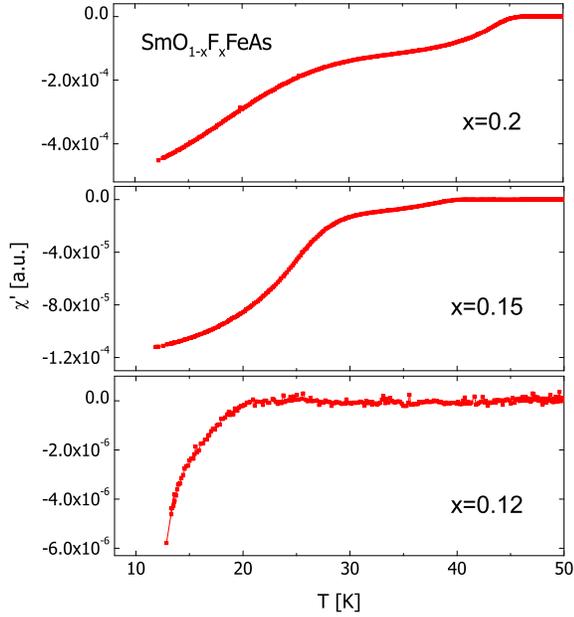}}
   \caption{The low temperature magnetic AC susceptibility of the SmFeAsO$_{1-x}$F$_{x}$ samples with different doping.}
   \label{FigSusc}
\end{figure}

In Fig.\ref{FigResist}, the resistivity curves are given that
show the critical temperature of about 20 K, 39 K and 45 K for
$x$ = 0.12, 0.15 and 0.2 respectively. As expected, the doping
also increases the sample metallicity. Magnetic AC
susceptibility on the same specimens (Fig.\ref{FigSusc}) shows the
superconducting transition at approximately the same temperatures
as in Fig.\ref{FigResist}, however, the curves for the two highest
doping levels show a "double-transition" structure
due to the granular nature of these samples \cite{SenatoreCM08}.

\begin{figure}[thb]
   \centerline{\includegraphics[width=\linewidth,clip=true]{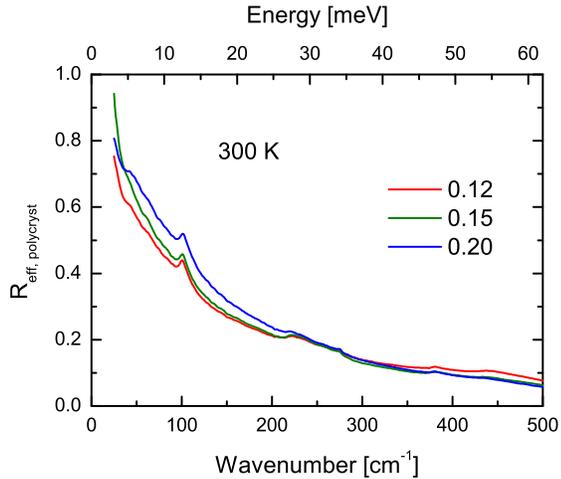}}
   \caption{Reflection coefficient of polycrystalline unpolished samples of SmFeAsO$_{1-x}$F$_{x}$
   with doping levels $x$ = 0.12, 0.15 and 0.2. }
   \label{FigRefl}
\end{figure}

The far infrared reflectance coefficient $R$ of the three
samples at room temperature obtained using a gold mirror as a
reference are shown in Fig. \ref{FigRefl}. The reflectivity is
quite low due to a combination of two factors: First, the specular reflection from a rough surface falls rapidly as a function of decreasing wavelength due to diffuse scattering. In addition, in view of the structural similarities to the cuprates, the optical anisotropy is expected to be high, and the metallic screening along the c-axis is expected to be much weaker than along the planes. The reflection coefficient becomes an effective medium average of both crystallographic directions, intermediate between the ab-plane and the c-axis response. In view of the fact that it is impossible to obtain separate ab-plane and c-axis tensor components from the optical response of a polycrystalline sample, we will concentrate here on the reflectivity spectra and the influence of the superconducting phase transition on those spectra.

\begin{figure}[thb]
   \centerline{\includegraphics[width=\linewidth,clip=true]{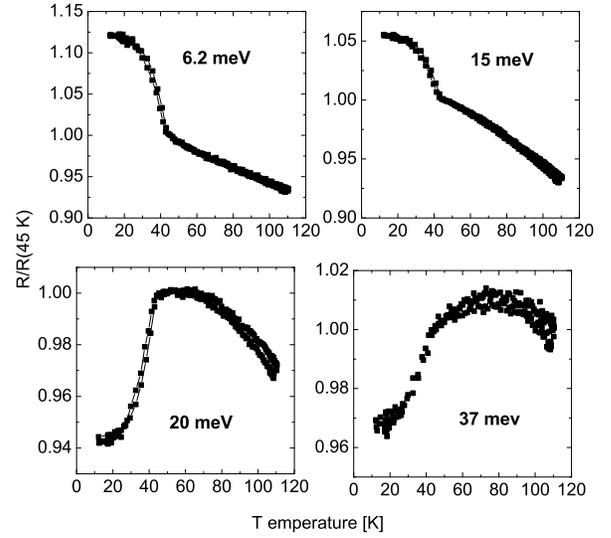}}
   \caption{Temperature dependence of the reflectivity ratio at 4 different energies for SmFeAsO$_{0.8}$F$_{0.2}$.}
   \label{FigTdep}
\end{figure}

At ambient temperature all reflection coefficients exhibit the -for a metal- characteristic Hagen Rubens behaviour
($1-R \propto \omega^{1/2}$) for  $\omega\rightarrow 0$. In this limit the reflectivity increases with increasing fluor concentration.  The features at 100 cm$^{-1}$ (12 meV), 220 - 270 cm$^{-1}$ (27 - 34 meV), 380 cm$^{-1}$ (47 meV) and 440 cm$^{-1}$ (55 meV) are probably all due to phonons with their polarization either along or perpendicular to the planes. The positions of these peaks confirm the infrared spectroscopic data on (Nd,Sm)FeAsO$_{0.82}$F$_{0.18}$ \cite{DubrokaCM08}, LaFeAsO$_{0.9}$F$_{0.1}$ \cite{DrechslerCM08} and CeFeAsO \cite{ChenCM08}.

\begin{figure}[thb]
   \centerline{\includegraphics[width=\linewidth,clip=true]{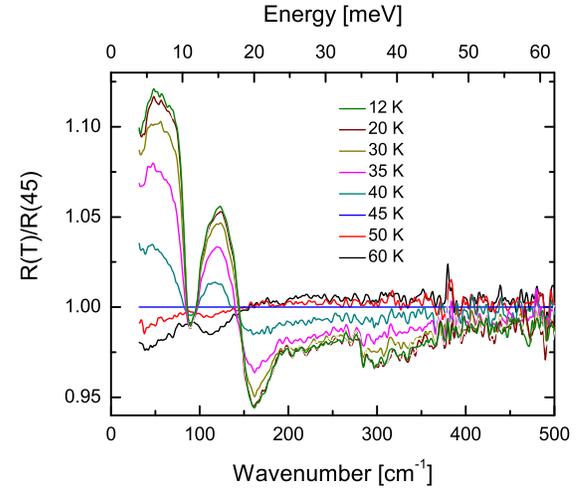}}
   \caption{Reflectivity ratio for the sample with $x$=0.2 for different temperatures. All reflectivity graphs are divided by the reflectivity at 45 K. }
   \label{FigReflRatio}
\end{figure}

The effect of the superconducting phase transition on the infrared response is revealed by the temperature dependence of the
reflectivity for four different photon energies (6.2, 15, 20 and 37 meV), presented in Fig.\ref{FigTdep}. A sharp upward kink at $T_{c}$ is
observed at low energies. The curve at higher energies also show a clear kink in the opposite direction. The observation of large reflectivity changes associated with the superconducting phase transition justifies us to concentrate in the remaining discussion on the reflectivity ratios $R(T)/R(T_n)$ normalized by the reflectivity at a temperature $T_n$ just above $T_c$.

\begin{figure}[thb]
   \centerline{\includegraphics[width=\linewidth,clip=true]{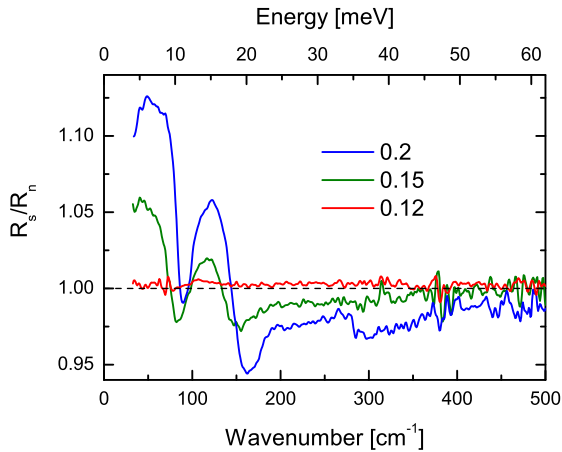}}
   \caption{Reflectivity ratio $R_{s}/R_{n}$ for samples with doping levels of 0.12, 0.15 and 0.2. $R_{s}$ is taken at 12 K and $R_{n}$ is the reflectivity at a temperature just above $T_{c}$.}
   \label{FigDopDep}
\end{figure}

This ratio, shown in Fig.\ref{FigReflRatio} for the $x=0.2$ sample, has
a rich structure as a function of frequency: In the
superconducting state, the reflectivity ratio increases below
18 meV reaching a maximum value of 1.12 at about 7 meV and
lowers at higher photon energies, showing a pronounced dip at
20 meV. This behavior agrees with the Mattis-Bardeen model for
a BCS superconductor, but would also be consistent with the
appearance of a strong plasma mode due to the condensate. As in
the earlier infrared experiments on polycrystalline samples of
the high-$T_{c}$ cuprates \cite{TannerTimusk} it is premature
to distinguish between these two possibilities. In the first
scenario, the value of the 2$\Delta$ can be very roughly
estimated to be between 10 and 20 meV, which would not be
inconsistent with the weak-coupling BCS gap ratio, observed in
a recent point-contact tunneling measurement
\cite{ChenNature08}.

Several additional spectral structures, in particular the prominent dip at about 11 meV, reveal the presence of infrared optical phonons, the strength in the reflectivity signal of which changes when the sample becomes superconducting. The dip in the reflectivity ratio reveals a temperature evolution whereby the peak seen at room temperature at 12 meV transforms into a minimum at 11 meV at low temperature. This is the expected behaviour when metallic screening of the phonon is very weak or absent above T$_c$, and forms only in the superconducting state. Since the conduction along the ab-plane is metallic, we attribute the 11 meV dip in the reflectance ratio to a c-axis optical phonon, and we conclude that the conductivity along the c-axis has an evolution as a function of temperature which is strongly similar to the cuprates, i.e. incoherent above T$_c$, and coherent below T$_c$\cite{tsvetkov98}.
The weaker structure between 25 and 35 meV is probably also due to phonon modes. It is less likely that the dip at 20 meV
is due to a lattice vibration given the absence of any structure in the absolute reflectivity spectra in this range.
%In view of similarities to the cuprates we speculate that this feature reveals the c-axis Josephson plasmon\cite{tsvetkov98}.

Fig. \ref{FigDopDep} shows the reflectivity ratio for three
different doping levels. Comparing to the
highest doping of 0.2, we see that for $x$= 0.15 and 0.12 the
low energy structures have strongly reduced in intensity and
also have slightly shifted to lower frequencies. For $x$ =
0.12, almost no effect of the superconducting transition is
discernible.

\begin{figure}[thb]
   \centerline{\includegraphics[width=\linewidth,clip=true]{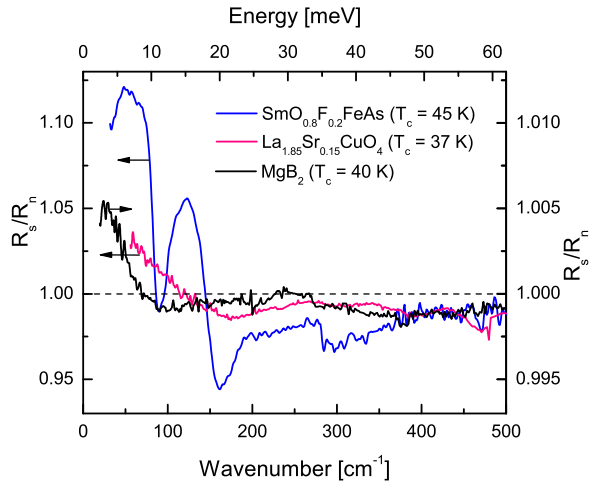}}
   \caption{The spectra of $R_{s}/R_{n}$ for three different superconductors with similar values of $T_{c}$.}
   \label{FigThreeCompounds}
\end{figure}

We compare the superconducting-to-normal
state reflectance ratio spectra for polycrystalline samples
from different families of superconductors with
$T_{c}$ in the same temperature range. Fig. \ref{FigThreeCompounds} presents
$R_{s}/R_{n}$ of SmFeAsO$_{0.8}$F$_{0.2}$ ($T_{c}$ = 45 K),
La$_{1.85}$Sr$_{0.15}$CuO$_{4}$ ($T_{c}$ = 37 K)
\cite{KuzmenkoPRL03} and MgB$_{2}$ ($T_{c}$ = 40 K)
\cite{KuzmenkoSSC02}. The
La$_{1.85}$Sr$_{0.15}$CuO$_{4}$ spectra are an effective medium\cite{StroudPRB75} simulation taking the experimental ab-plane and c-axis dielectric function measured of single crystal La$_{1.85}$Sr$_{0.15}$CuO$_{4}$ as input parameters\cite{KuzmenkoPRL03}.

The comparison between these three reveals, that the spectral range where the superconductivity induced changes of SmFeAsO$_{0.8}$F$_{0.2}$ and La$_{1.85}$Sr$_{0.15}$CuO$_{4}$ are in the range of 25 meV (about 9 times k$_{B}$T$_{c}$), whereas for MgB$_{2}$ this only 15 meV (4 times k$_{B}$T$_{c}$). While this energy does in general not correspond directly to the gap value, the similarities between SmFeAsO$_{0.8}$F$_{0.2}$ and La$_{1.85}$Sr$_{0.15}$CuO$_{4}$, sets them apart from the electron-phonon mediated superconductor MgB$_{2}$.

In summary, we have measured far-infrared reflectivity spectra
of the new superconductor SmO$_{1-x}$F$_{x}$FeAs ($x$ = 0.12,
0.15 and 0.2) in the normal and superconducting states. The
temperature dependent reflectivity of polycrystalline samples
shows a sharp superconducting anomaly. It strongly increases
for the photon energy smaller than 18 meV and decreases at
higher energies. However, the shape of the
superconducting-to-normal state reflectivity ratio is also
affected by an optical phonon at about 11 meV. The energy scale
where superconductivity has an noticeable effect on infrared
spectra is quite broad, comparable the one in high T$_{c}$
cuprates and much larger than in MgB$_{2}$. Although the
observed spectra are consistent with a BCS superconductivity,
at this stage we cannot reliably extract the value of the gap
given the strong anisotropy and a possible presence of the
c-axis Josephson plasmon (as observed in cuprates). Future
measurements on single crystals are needed to obtain quantitative results on, for example, the optical conductivity tensor of these strongly anisotropic materials.

This work is supported by the Swiss National Science Foundation through Grant No. 200020-113293 and the National Center of Competence in Research (NCCR)``. Materials with Novel Electronic Properties-MaNEP''. We gratefully acknowledge fruitful discussions with R. Fl\"ukiger and D. Scalapino.

\end{document}